\pdfoutput=1
\documentclass[twoside,slac_one]{revtex4}
\usepackage{ifpdf}
\ifpdf
  
  \usepackage{microtype}
\fi
\usepackage{hyperref}
\usepackage{graphicx}
\usepackage{fancyhdr}
\usepackage{amsmath} 
\usepackage{bm}
\usepackage{amsxtra}
\usepackage{amssymb}
\usepackage{amsthm}
\usepackage{latexsym}
\usepackage{lscape}

\usepackage{xspace}
\usepackage{booktabs}
\RequirePackage{subfigure}

\newcommand{\dzero}      {D0\xspace}
\def\et{\ensuremath{E_{T}}}
\def\met{\ensuremath{\not\!\!\et }}
\def\vecmet{\not\!\!\vec\et }
\newcommand{\pt}{\ensuremath{p_{_T}}\xspace}
\newcommand{\ptone}{\ensuremath{p_{_{T1}}}\xspace}
\newcommand{\pttwo}{\ensuremath{p_{_{T2}}}\xspace}
\newcommand{\ptonetwo}{\ensuremath{p_{_{T1,2}}}\xspace}

\makeatletter
\@ifpackageloaded{hyperref}{%
  \catcode`\Q=3
  \newcommand\trim@spaces[1]{%
    \romannumeral-`\q\trim@trim@\noexpand#1Q Q%
  }
  \long\def\trim@trim@#1 Q{\trim@trim@@#1Q}
  \long\def\trim@trim@@#1Q#2{#1}
  \catcode`\Q=11 
  \newcommand{\eprint}[1]{\href{http://arxiv.org/abs/\trim@spaces{#1}}{\trim@spaces{#1}}}
}%
{\newcommand{\eprint}[1]{#1}}
\makeatother

\begin{document}

\title{Search for Universal Extra Dimensions with the \dzero Experiment}

%

\author{J. D. Mansour on behalf of the \dzero Collaboration}
\affiliation{ II. Physikalisches Institut, Georg-August-Universit\"at G\"ottingen, Germany}

\begin{abstract}
A search for signs of universal extra dimensions (UED) has been performed with the \dzero experiment, using events with two same-sign muons.  The considered minimal UED model includes one extra dimension, and has a stable lightest Kaluza-Klein particle (LKP) which is a dark matter candidate.  In the search, 7.3~fb$^{-1}$ of \dzero data, collected in $p\bar p$ collisions at the Fermilab Tevatron collider at $\sqrt{s}=1.96$~TeV, have been used.
\end{abstract}

\makeatletter
\hypersetup{pdfauthor={Jason D. Mansour},pdftitle={\@title}}
\makeatother

\maketitle

\thispagestyle{fancy}


\section{Theory}
In the past century, many theories have been suggested that include extra spatial dimensions.  An early attempt was made by Kaluza~\cite{kaluza} to unify electromagnetism with general relativity by introducing a fifth dimension.  This theory was extended by Klein~\cite{klein}, who argued that the additional dimension is not observed because it is compactified, or ``rolled up''.  This branch of research was later abandoned, since as a classic theory it did not account for nuclear interactions, and it seemed to suggest the existence of a new particle that was not observed.

Interest in theories with extra dimensions has increased again in the past decade, since they promise solutions to some outstanding problems in physics, and make predictions that can be tested in collider experiments such as the Tevatron or the LHC.  One theory that might hold a solution to the hierarchy problem is the Randall-Sundrum model~\cite{rs}, which assumes a five-dimensional spacetime.  The standard model particles are confined to a four-dimensional brane, while gravity can propagate outside into the extra dimension, or bulk.  The model of large extra dimensions by Arkani-Hamed, Dimopoulos and Dvali (ADD)~\cite{add,add2} shares the idea of gravity being able to propagate into an extra dimension, and aims to explain why it is so weak compared to other forces.

In models with universal extra dimensions (UED)~\cite{ued1}, not only gravity, but all forces are allowed to propagate into the bulk.  The movement of particles in the extra dimension(s) is not seen as such, but since the particles have additional kinetic energy, they seem to be copies of standard model particles with a higher rest mass.  In the following, the minimal UED model~\cite{ued2} with one compactified extra dimension will be assumed.  This model has two parameters, the compactification scale $R^{-1}$, and a cutoff scale $\Lambda$ up to which the effective theory is valid.  Here, $\Lambda=10000$~GeV is assumed.  The compactification introduces periodic boundary conditions, so that the momentum in the extra dimension can only take discrete values.  This leads to the appearance of a series of excitations of SM particles, also called a Kaluza-Klein (KK) tower.  Generally, only the first excitations ($n=1$) can be seen at current collider experiments.  The masses of the KK particles are given to first order (without radiative corrections) by $M_n^2 = M_0^2 + n^2/R^2$, where $M_0$ is the SM mass.  Since momentum is conserved separately in the extra dimension, so are these excitations, and one can introduce a new conserved quantity called KK parity.  This multiplicative quantum number, similar to $R$ parity in certain supersymmetric models, ensures that KK particles can only be produced in pairs, and furthermore that the lightest KK particle (LKP) is stable, making it a good candidate for cold dark matter.

In the considered model, the excited photon ($\gamma_1$) is the LKP.  The next heavier particles are KK-leptons ($\ell_1$), gauge bosons ($Z_1/W_1$), quarks (SU(2) doublet $Q_1$ or singlet $q_1$), and finally excited gluons ($g_1$).

\begin{figure}[ht]
\centering
\subfigure[]{
\includegraphics[scale=0.80]{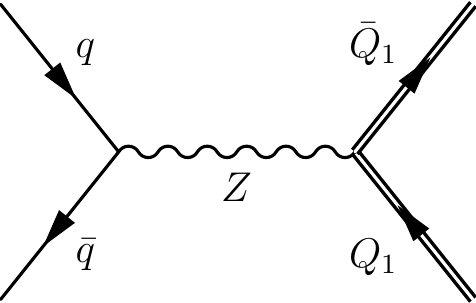} \label{fig:feynman_a}}
$\qquad$
\subfigure[]{
\includegraphics[scale=0.80]{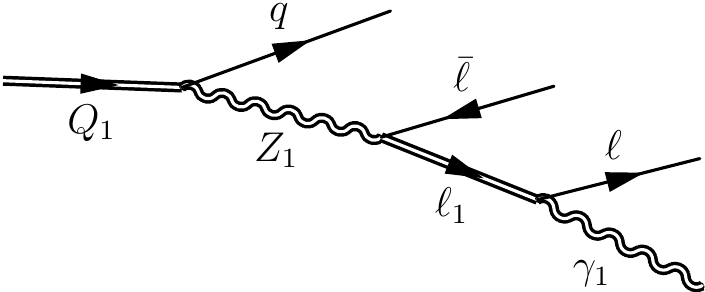} \label{fig:feynman_b}}
\caption{(a) Production of pair of KK-excited quarks ($Q_1\bar Q_1$) via $Z$ boson.
(b) Decay of a KK quark into a jet, two opposite sign leptons, and an LKP.  The double lines indicate KK particles.} \label{feynman}
\end{figure}

The production mode of KK particles at the Tevatron is via pair production of KK gluons or quarks.  These decay in cascades into standard model particles and LKPs.  A typical production and subsequent decay is shown in Fig.~\ref{feynman}.  The $Q_1$ decays into an excited gauge boson ($Z_1$ or $W_1$) and a standard model quark.  The excited boson decays further into a standard model lepton and a KK-lepton, which finally decays into another SM lepton and an LKP ($\gamma_1$).  In the final state, there are up to four charged leptons, two jets, and missing transverse energy (\met)\ (from LKPs and possibly neutrinos).  Since the emitted SM leptons may have a very low \pt when the KK particles have similar masses, and thus might be below reconstruction threshold, only the presence of two leptons is demanded.  To suppress SM background from $Z\rightarrow \mu\mu$, both leptons must have the same sign.  In this analysis, the like-sign dimuon case will be considered, since muons have the highest reconstruction efficiencies of the charged leptons.

Recently, searches have been conducted in the diphoton channel by \dzero~\cite{dzeroDiphoton} and ATLAS~\cite{atlasDiphoton} for a related, but different model, in which the lightest KK particle $\gamma_1$ may decay into a graviton and a photon.  However, for the minimal UED model studied here, in which the LKP is stable, there have been no direct limits to date.

\section{Experiment}
The Tevatron is a proton-antiproton collider with a center-of-mass energy of $\sqrt{s} = 1.96$~TeV.  In Run II, from 2001 through Sep. 2011, it delivered more than 10~fb$^{-1}$ to each of its two experiments, \dzero and CDF.  The \dzero detector~\cite{dzero} is a general purpose detector with nearly $4\pi$ coverage.  The innermost system is the silicon microstrip tracker (SMT), surrounded by the central fiber tracker (CFT) which utilizes scintillating fibers in stereo configuration.  These systems form the central tracker.  Outside of the tracking system is the LAr/Ur sampling calorimeter, which is split into central and endcap regions.  Beyond the calorimeter is a dedicated system for triggering on and detecting muons, consisting of mini drift tubes (MDTs) and proportional drift tubes (PDTs).  The \dzero detector utilizes toroidal and solenoidal magnets of each nearly 2~T to curve the trajectories of charged particles in the tracker or the muon system, respectively.


\section{Samples}
In this analysis, 7.3~fb$^{-1}$ of \dzero data are used, collected in Run II until July 2010.  All events which fired a single muon trigger and pass data quality requirements are used.

Most standard model backgrounds are modeled with Monte Carlo.  $W\,+\,$jets, $Z\,+\,$jets and $t\bar t$ events are generated with \textsc{alpgen}~\cite{alpgen} interfaced to \textsc{pythia}~\cite{pythia6} for hadronization and showering, while diboson events ($WW$, $WZ$ and $ZZ$) are simulated using \textsc{pythia} alone.  In both cases, the CTEQ6L1 parton distribution functions~\cite{CTEQ6L1} were used.  The $W/Z$ + jets and diboson samples were scaled to a next-to-leading-order (NLO) prediction from \textsc{MCFM}~\cite{Campbell:2002tg,Campbell:2003hd}.  Cross sections for $t\bar t$ in next-to-next-to-leading-order (NNLO) were obtained from~\cite{ttCrossSection}.  Signal MC was generated for 9 evenly spaced points from $R^{-1} = 200$~GeV to 320~GeV, using \textsc{pythia} 6.421 and CTEQ5L PDFs~\cite{CTEQ5}.

All MC samples are processed through a full \textsc{geant}~\cite{geant} based detector simulation, and events are reconstructed using the same algorithms that are applied to data.  Several reweightings are applied to MC to improve agreement with data.  MC is reweighted so its instantaneous luminosity distribution matches the luminosity profiles of data.  In the $Z$+jets sample, a correction is applied to the \pt distribution of the $Z$ boson.  Finally, reweightings are applied to to account for differences in trigger and object identification efficiencies.

\subsection{Preselection}
As a preselection, events with two same-sign muons which fulfill certain reconstruction quality criteria are selected.  Muons must have a transverse momentum $\pt > 5$~GeV$/c$.  Hits in the muon system are matched to central tracks, and the global fit of the muon track must have a $\chi^2 < 9.5$.  There must be at least two CFT hits.  A cosmic veto is applied, where the time between scintillator hits must be less than 10~ns.  The distance of closest approach (dca) of the muon track to the primary vertex must be less than 0.04~cm, or if there is no SMT hit, less than 0.2~cm.  Finally, a cut $|z(\mu_1) - z(\mu_2)| < 1$~cm is applied, where $z$ is the distance between the muon and the primary vertex along the beamline.

For modeling and suppression of the multijet background, muon isolation is an important quantity. Calorimeter isolation $I_\mathrm{cal}$ is defined as the sum of $E_T$ in the calorimeter in a hollow cone $0.1 < \Delta R < 0.4$ around the muon, divided by the muon $\pt\cdot c$.  Here, $\Delta R$ is a measure of angular distance, given by $\Delta R = \sqrt{\Delta \eta^2 + \Delta \varphi^2}$.  Additionally, track isolation $I_\mathrm{trk}$ is defined as the muon of $\pt$ of all tracks in the cone $\Delta R < 0.5$ around the muon, divided by its $\pt$. With these isolation variables, the muons are separated into three categories: A tight muon (T) has $I_\mathrm{cal} < 0.4$ and $I_\mathrm{trk} < 0.12$. A loose muon (L) is not tight, and has $I_\mathrm{cal} < 0.4$ and $I_\mathrm{trk} < 0.25$. A non-isolated muon (N) is neither tight nor loose. 

From the preselection, a main selection sample $S$ is formed, containing all events where one muon of the same-sign pair is tight, and one is tight or loose (TT or TL).  Cuts are applied on $\Delta \varphi$ between the two muons:  Events with $\Delta \varphi < 0.25$ are removed, and to cut away some of the multijet background, also back-to-back events ($\Delta \varphi > 2.9$).  Only events with both muons in the central region $|\eta_\mathrm{det}| < 1.5$ are used, since the tracking efficiency is much lower and less well modeled outside this region.  Finally, a minimum $\pt$ cut of $\pt > 10$~GeV$/c$ is applied.

\section{Background estimation}
While most backgrounds are modeled with MC, for the estimation of the multijet and charge misidentification backgrounds, data-driven techniques were used.  The modeling of these backgrounds shall be described in the following.

\subsection{Multijet}
The QCD multijet background is mostly made up of muons emitted from jets from $b\bar b/c\bar c$ decays.  One characteristic of this background is that, since jets tend to be balanced, the muon pairs are mostly back to back, whereas the angles of signal muons are much less correlated since they come from cascade decays.  Since multijet muons originate in jets, they tend to be less isolated than signal muons.  However, secondary muons from jets can become isolated and be misidentified as prompt muons.  The probability of this is higher for lower $\pt$ muons, and so the multijet background is larger in that region.  These properties can be used to model the multijet background from data.

A multijet-enriched sample $Q$ is defined with the same criteria as $S$, but the following exception:  the first muon is tight as in the $S$ sample, but the second muon is non-isolated (TN).  This sample provides the shape of the multijet background, while the normalization is calculated in an orthogonal region where the multijet background dominates.  This is done by taking the ratio of integrals of the $S$ and $Q$ samples, while flipping the $\pt$ cut for the most isolated muon (lowest $I_\mathrm{cal}$): 
\[
   N_i = \frac{\int S}{\int Q}\,, \quad
   i = 0, 1, \geq 2
\]
The resulting normalization factor is strongly dependent on the jet multiplicity $i$, thus different normalizations $N_i$ are determined for zero, one, and two or more jets.  These factors are applied as weights to the $Q$ sample (in the region $\pt > 10$~GeV).
 
\begin{figure}[!hbt]
  \centering
  \includegraphics[width=9.0cm]{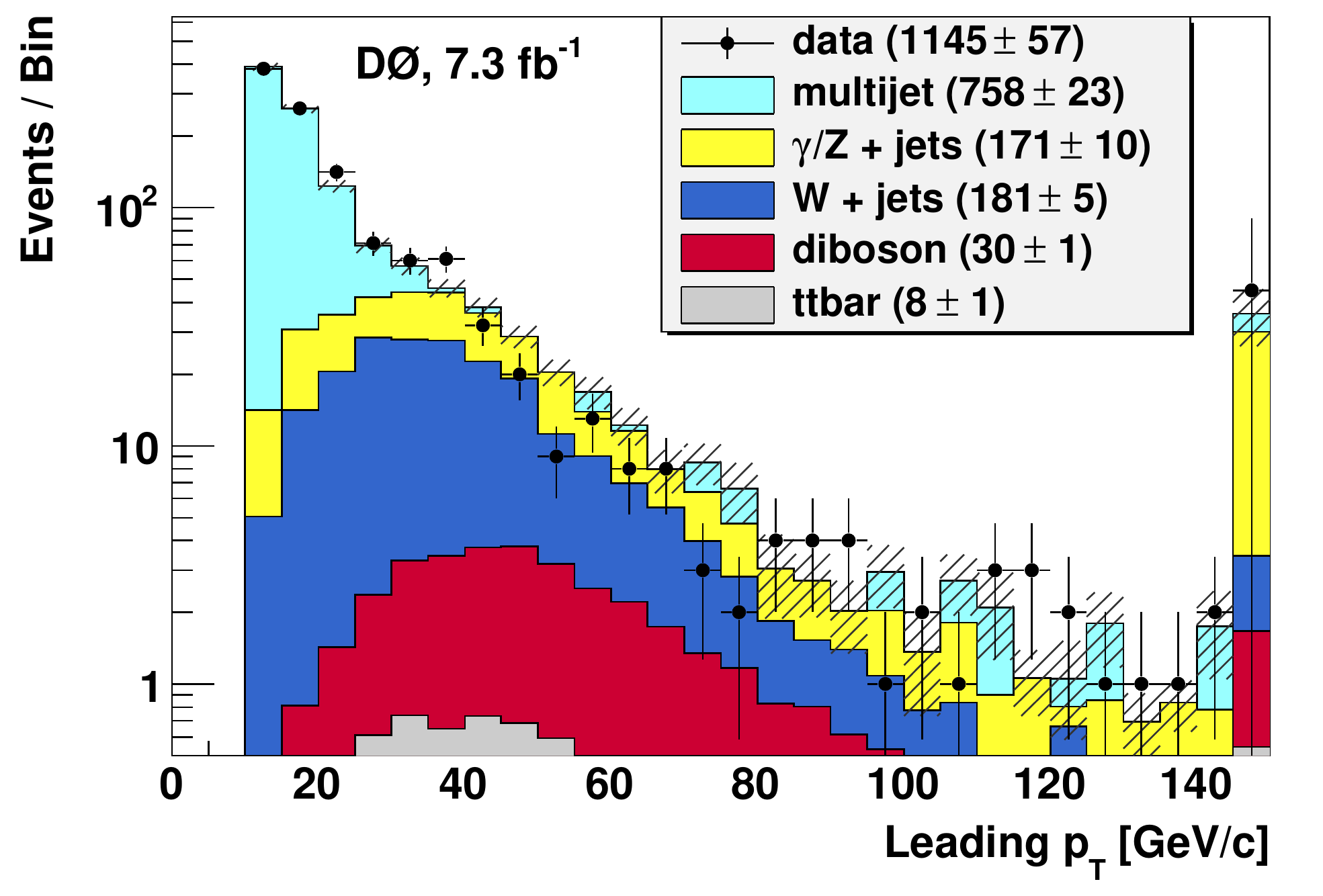}
  \caption{Comparison of data and background estimations for the leading muon $\pt$ distribution.}
  \label{fig:backgrounds}
\end{figure}

While the assumption that the multijet background is dominant in the $Q$ sample is valid for low $\pt$, there is a significant contribution of electroweak (EW) processes, mainly $W$+jets, in the higher $\pt$ region.  Since these events are already considered in MC, this leads to double counting.  This is solved by estimating this EW contamination, and subtracting it from the reweighted $Q$ sample.  The EW contamination in $Q$ is estimated by applying the $Q$ selection and the $N_i$ weights to EW MC.  The additional physical constraint that the multijet background goes to zero (and must not be negative) when $\pt(\mu) \rightarrow \infty$ is used to derive a normalization factor $f$ for the subtracted EW contamination.  The final estimation for the background is thus given by:
\[
  \mathrm{BG}_i = Q^\mathrm{data} \cdot N_i - f \cdot Q^\mathrm{MC} \cdot N_i + S^\mathrm{MC} \,, \quad i = 0,1,\geq 2 \,.
\]
Here, the first two terms give the final multijet estimation.  $Q^\mathrm{data}$ and $Q^\mathrm{MC}$ denote the $Q$ selection applied to data and MC, respectively, and $S^\mathrm{MC}$ is the estimation for non-multijet backgrounds from MC ($W/Z$+jets, diboson, $t\bar t$).   
The leading $\pt$ distribution including the multijet background is shown in Fig.~\ref{fig:backgrounds}.

\subsection{Charge mismeasurement}

A sizable background is due to $Z\rightarrow \mu^+\mu^-$ opposite-sign events, where the charge of one muon gets mismeasured.  This can especially happen at high $\pt$ when the track of the muon is almost straight, or when additional hits from other charged particles or noise are present near the track.  A small charge mismeasurement probability can lead to a large background in the same-sign channel because of the large number of events in the $Z$ peak.  In principle, this charge flip (CF) background is taken into account by running the $Z$+jets MC through the detector simulation.  However, since charge mismeasurement is not well simulated, it is advisable to confirm the MC prediction with an independent estimation, which is determined using a data driven method~\cite{chargeflip}.
 
The idea behind this method is that the \dzero detector provides two uncorrelated muon charge measurements: One ``local'' measurement from the track in the muon chambers, and one from the matched track in the central tracker.  Since the accuracy of the central tracker is much larger, this measurement is usually meant when simply referred to the ``charge'' of a reconstructed muon.  

Data events are split into three categories, depending on how often the local and tracker measurements agree: AA if the measurements agree for both muons, AD if there is a disagreement for one muon in the pair, and DD if both muons have disagreeing local and track charges.  The number of muons in each category is then given by:
\begin{eqnarray}
  N_\mathrm{AA} = P^\mathrm{true}_\mathrm{AA} \cdot N_\mathrm{true}
                              + P^\mathrm{flip}_\mathrm{AA} \cdot N_\mathrm{flip} \nonumber\\
  N_\mathrm{AD} = P^\mathrm{true}_\mathrm{AD} \cdot N_\mathrm{true}
                              + P^\mathrm{flip}_\mathrm{AD} \cdot N_\mathrm{flip} \label{eqn:chargeflip} \\
  N_\mathrm{DD} = P^\mathrm{true}_\mathrm{DD} \cdot N_\mathrm{true}
                              + P^\mathrm{flip}_\mathrm{DD} \cdot N_\mathrm{flip} \nonumber
\end{eqnarray}
Here, $P^\mathrm{true}_\mathrm{AA} = \varepsilon_\mathrm{loc} \cdot \varepsilon_\mathrm{loc}$ is the probability that in a true like-sign event the measurements agree for both muons (AA).  Since the tracker is assumed to be much more accurate than the muon system, its inefficiency can be neglected, and the probability is the efficiency of the local measurement $\varepsilon_\mathrm{loc}$ squared.  $P^\mathrm{flip}_\mathrm{AA}$ is the respective probability for an opposite-sign event.  Since all considered events are in the like-sign sample, the tracker measurement of one of the muons must be wrong, and since it is an AA event, so must be the local measurement.  It follows $P^\mathrm{flip}_\mathrm{AA} = \varepsilon_\mathrm{loc} \cdot (1 - \varepsilon_\mathrm{loc})$.  Similar expressions can be found for the other probabilities.  The fraction of charge flip events within the like-sign sample is given by
\[
  f_\mathrm{flip} = N_\mathrm{flip} / (N_\mathrm{flip} + N_\mathrm{true}) \,,
\]   
and can be obtained by solving the system of equations~\eqref{eqn:chargeflip}.  This fraction multiplied by the total number of data events yields an estimation of the charge flip background.  The obtained number, $161.7 \pm 32.4$, is in agreement with the number of events from the $Z\rightarrow \mu\mu$ background MC, $170.6$.  From the difference between both, a systematic uncertainty for the CF background is calculated.  Since that difference is small compared to the statistical uncertainty of the CF estimation, both are added in quadrature to arrive at a systematic uncertainty of 20.8\%.

\section{Signal extraction}

A multivariate analysis (MVA) is used to take advantage of correlations between variables~\cite{TMVA}.  The following variables are uses as inputs to the MVA:

\begin{itemize}
 \item Leading and second transverse momentum (\ptone and \pttwo)
 \item Missing transverse energy ($\met$)
 \item Dimuon invariant mass ($M_\mathrm{pair}$)
 \item Angle between the two muons in the transverse plane ($\Delta\varphi$)
 \item Fit $\chi^2$ of the track matched to the first or second muon (the background contains many events with mismeasured tracks)
 \item Number of Jets ($N_\mathrm{jets}$)
 \item Product of missing transverse energy and second $\pt$ ($\met \cdot \pttwo$), since in the multijet background either one of them tends to be low
 \item Topological variables $M_{T1}$ and $M_{T2}$, calculated as the transverse mass between $\met$ and one of the muons:
 \[
   M_{T1,2} := \sqrt{2 \met \cdot \ptonetwo (1-\cos\Delta\varphi(\vecmet,\mu_{1,2}))}
 \]
 These provide a good separation between different types of background.
\end{itemize}

Some additional cuts are applied before application of the MVA to reduce specific backgrounds or not well-modeled regions: $\met > 25$~GeV to reduce the multijet background, $\ptone = 10 \ldots 200\;\mathrm{GeV}/c$, $\pttwo > 10\;\mathrm{GeV}/c$, a cut on the dimuon invariant mass $M_\mathrm{pair} = 20\ldots250\;\mathrm{GeV}/c^2$, and finally $|\mathrm{dca}_{1,2}| < 0.05\;$cm to reduce pile-up and badly measured tracks.

\begin{figure}[!hbt]
  \centering
  \subfigure{\label{fig:pT1}}
  \subfigure{\label{fig:pT2}}
  \subfigure{\label{fig:MET}}
  \subfigure{\label{fig:InvariantMass}}
  \includegraphics[width=12.0cm]{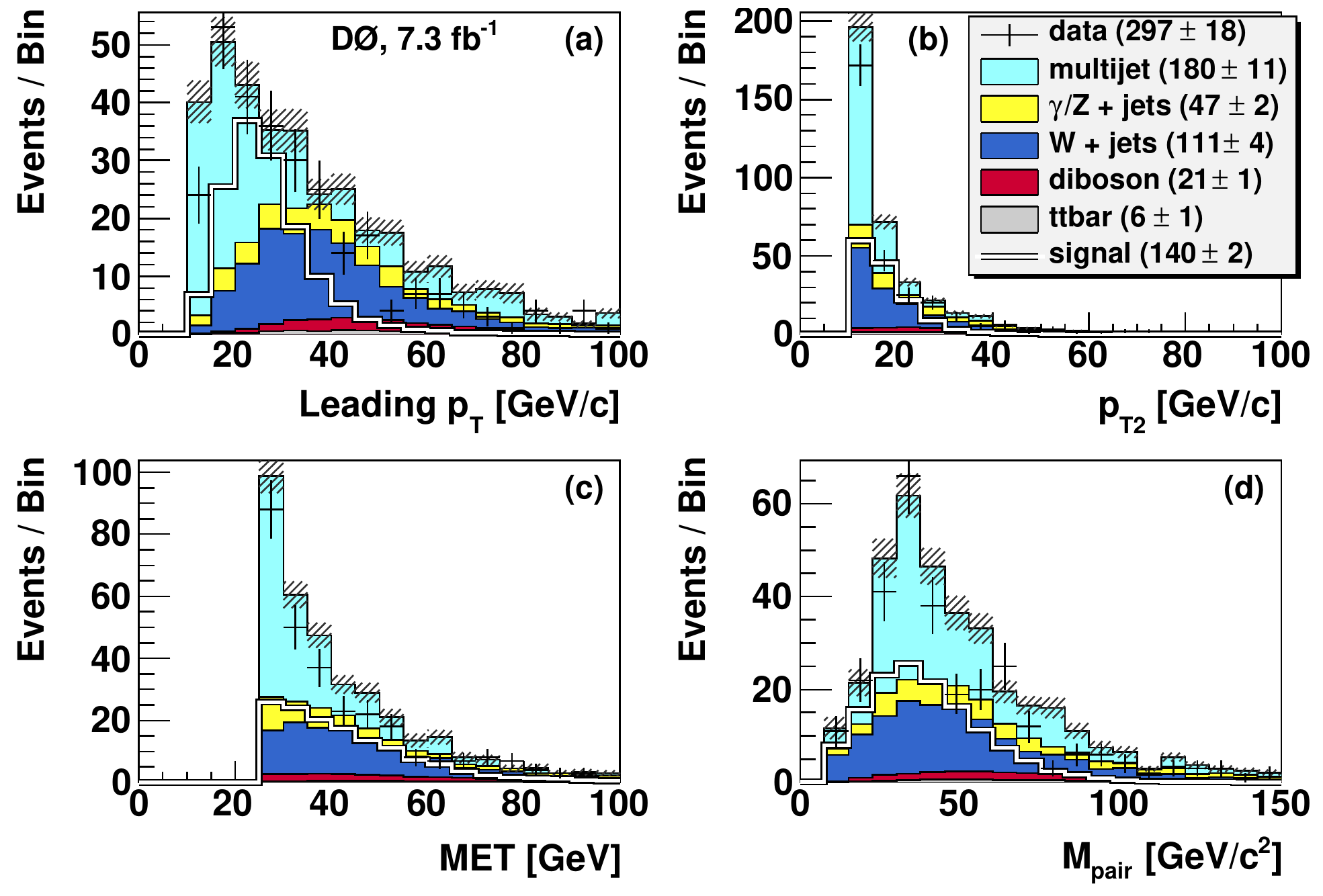}
  \caption[Some input variables for the multivariate analysis.]{Some input variables for the multivariate analysis. \subref{fig:pT1}~leading muon \pt, \subref{fig:pT2}~second muon \pt, \subref{fig:MET}~Missing transverse energy~$\met$ and~\subref{fig:InvariantMass} muon pair
invariant mass. The signal is shown for $R^{-1} = 245$~GeV.}
\label{fig:input_to_TMVA}
\end{figure}

\begin{figure}[!hbt]
  \centering
  \includegraphics[width=9.0cm]{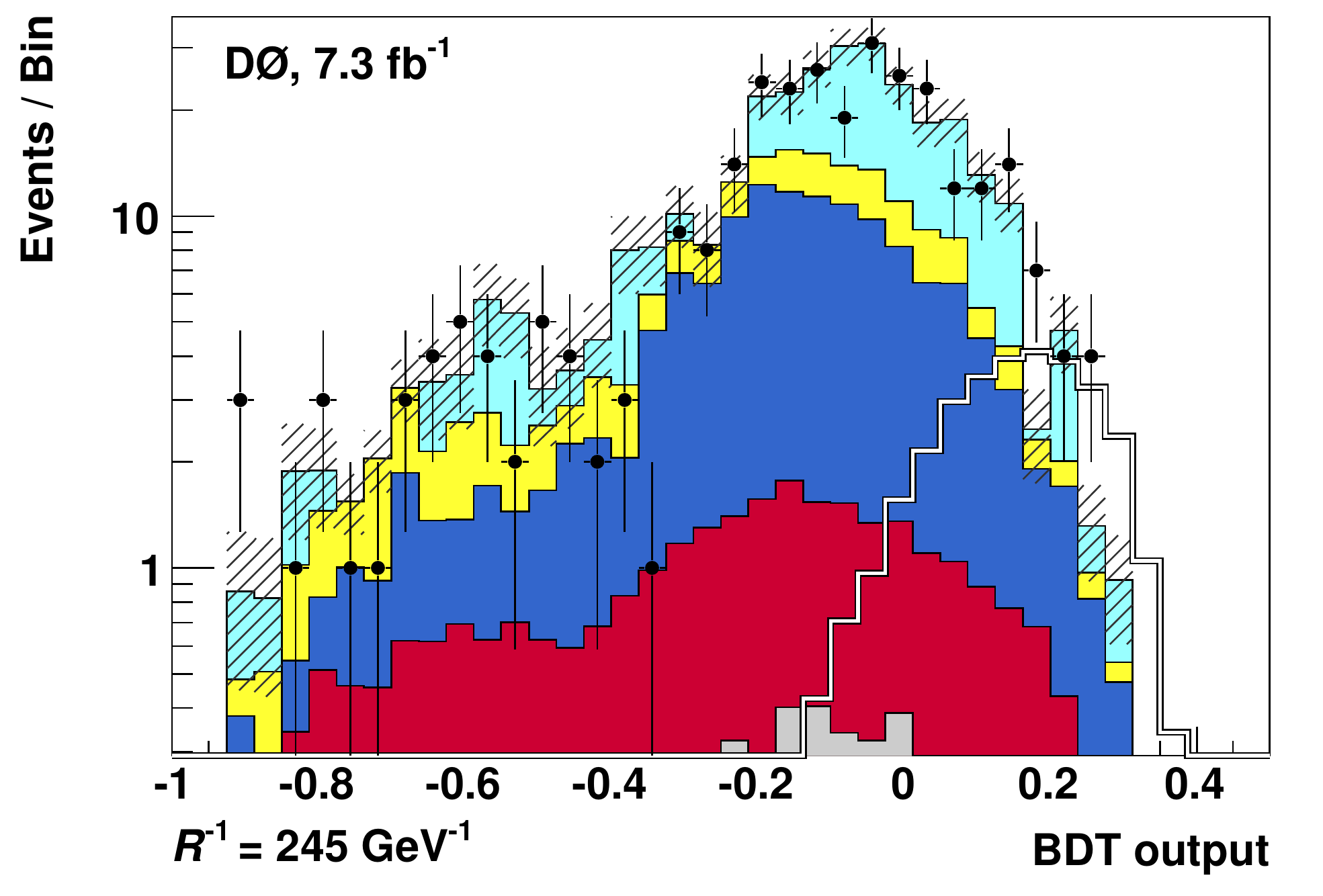} \\
  \caption{Output variable of the BDT for $R^{-1} = 245$~GeV.}
  \label{fig:tmva_output}
\end{figure}

A boosted decision tree (BDT) method was chosen as classifier.  Individual BDTs were trained for each signal point.  The output of the BDT for the signal point at $R^{-1} = 245\;$GeV can be seen in Fig.~\ref{fig:tmva_output}.  Higher values mean more signal-like events.  The separation of signal and background is clearly visible, and there is no significant excess of data compatible with a signal.

\begin{table}[ht]
\begin{center}
\caption{Sources of systematic uncertainties taken into account:}
\medskip
\begin{tabular}{lcr}
\toprule
\textbf{Systematic} & \textbf{Applied to}  & \textbf{Uncertainty}\\
\midrule
Luminosity		& All		& 6.1\% \\
PDF				& All		& 4.0\% \\
Jet energy scale	& All		& 4.1\% \\
Muon identification  & All		& 2.0\% \\
Tracking efficiency  & All		& 1.0\% \\
Isolation efficiency  & All		& 0.5\% \\
Trigger			& All	    	& 6.0\% \\
$Z/\gamma$ cross section  & $Z$+jets & 3.5\% \\
Charge flip     & $Z$+jets      & 20.8\% \\
$W$+jets cross section & $W$+jets & 8.5\% \\
Top pair production cross section & $t\bar t$ & 15.0\% \\
Diboson cross section & diboson & 7.0\% \\
Multijet estimation	& multijet & 40.3\% \\ 
\bottomrule
\end{tabular}
\label{tab:systematics}
\end{center}
\end{table}

\section{Conclusions}

A search for universal extra dimensions in the dimuon channel using 7.3~fb$^{-1}$ of \dzero data has been performed in the like-sign dimuon channel.  No excess of data over background was observed.  A publication of these results including limits on the mUED compactification scale is in preparation.

\begin{acknowledgments}
I would like to thank my colleagues Pedro Mercadante, Alexey Popov, Angelo Santos and Andrey Shchukin for the hard work they put into our analysis.
\end{acknowledgments}

\bigskip 

\end{document}